\documentclass[superscriptaddress,twocolumn,showpacs,prl,amsmath,amssymb]{revtex4}
\usepackage{graphicx}
\usepackage{bm}
\begin{document}
\title{Searching for a Supersolid in Cold Atom
  Optical Lattices}
\author{V.W. Scarola}
\affiliation{Condensed Matter Theory Center, 
Department of Physics, University of Maryland,
College Park, MD 20742}
\author{E. Demler}
\affiliation{Physics Department, Harvard University,
Cambridge, MA 02138}
\author{S. Das Sarma}
\affiliation{Condensed Matter Theory Center, 
Department of Physics, University of Maryland,
College Park, MD 20742}

\begin{abstract}
We suggest a technique for the observation of a predicted 
supersolid phase in extended Bose-Hubbard models which are 
potentially realizable in
cold atom optical lattice systems.  In particular, we discuss 
important subtleties arising from the existence of the trapping
potential which leads to an externally imposed (as opposed to
spontaneous) breaking of translational invariance.  We show, by 
carefully including the trapping potential in our theoretical
formalism, that noise correlations could prove 
instrumental in identifying the supersolid and density wave phases.  
We also find that the noise correlation peak 
width scales inversely with the relative size of trapped Mott domains.
\end{abstract}
\maketitle

Cold atom optical lattice systems are fast becoming ideal model
systems for studying quantum phases and quantum phase transitions in
various strong correlation condensed matter 
Hamiltonians \cite{Phillips}.  Experimental indicators 
for novel quantum collective phases such as the superfluid and the Mott 
insulator \cite{Greiner,Folling,Jaksch} 
already exist in cold atom systems, and there are 
concrete predictions under feasible and
realistic experimental conditions, for the existence of several other
interesting collective phases including various quantum spin 
phases \cite{Duan,Girvin}, 
fractional quantum Hall phases \cite{Sorensen}, and a strongly correlated
supersolid cold atom phase \cite{Scarola}.  

A key issue in the context of correlated cold atom quantum phases is
their actual experimental observability (in contrast to their possible
physical existence).  The experimental techniques (based mostly on 
interferometry) for studying cold atom optical lattices differ vastly
from the standard spectroscopic, thermodynamic, and transport
measurements prevalent in quantum condensed matter systems.
Therefore, it is not always manifestly obvious how to experimentally
observe and compellingly establish the various strongly correlated
collective phases in cold atom optical lattice systems even if their
existence is convincingly theoretically predicted.  The situation is
further complicated by the existence of the slowly varying externally
imposed confining trap potential which is usually present in 
cold atom systems.  In fact,
it is known that the confining trap potential introduces special
features which, if not accounted for carefully in the theory 
could lead to incorrect conclusions about the quantum 
phase diagram of the system.

We consider the experimental observability of the
predicted cold atom supersolid phase using the recently proposed 
\cite{Altman} and demonstrated \cite{Hadzibabic,Folling,Greiner2} 
noise correlation technique which is particularly suited to 
probing many-body states of cold atoms by directly investigating 
second order correlations in the absorption images of expanding
atomic gas clouds.  Conceptually, the noise correlation technique is
akin to solid state neutron, electron, or x-ray spectroscopy of
strongly correlated condensed matter systems where a suitable
correlation function is spectroscopically observed providing a direct
measure of strong correlation effects.  We show that the noise 
correlation function proves vital in distinguishing between spontaneous 
and externally broken translational symmetry in trapped cold 
atom optical lattice systems.    

We study the observable properties of 
two-dimensional cold atom, optical lattices  
in the presence of a parabolic trap modeled by 
the single-band, extended Bose-Hubbard Hamiltonian:
\begin{eqnarray}
H&=&-t\sum_{<i,j>}
\left(b^{\dagger}_{i}b^{\vphantom{\dagger}}_{j}
+h.c.\right)
+\frac{U}{2}\sum_i n_{i}(n_{i}-1)
\nonumber
\\
&+&V\sum_{<i,j>}
n_{i}n_{j}
-\sum_i \mu_{i} n_{i},
\label{Hex}
\end{eqnarray}
where $b^{\dagger}_{i}$ creates a boson at the site $i$ and 
$n_{i}=b^{\dagger}_{i}b^{\vphantom{\dagger}}_{i}$.  
The first term indicates the hopping, $t$, between
nearest neighbors, denoted by angular brackets.  The second term 
represents the on-site interaction energy cost, $U$.  The 
third term contains the
interaction energy cost, $V$, between nearest neighbors 
arising from the spatially
extended range of interaction possibly generated from a long-range
dipolar interaction \cite{Goral,Axel}, a short-range 
interaction between bosons in
higher bands of the lattice \cite{Scarola}, or a combination thereof.
A parabolic trapping potential of strength
$\kappa$ externally breaks translational invariance by modifying the chemical potential:
$\mu_i=\mu_0-\kappa\vert \bm{R}_i\vert^2$, 
where $\bm{R}_i=(i_x,i_y)$ for a two-dimensional square lattice.  We 
work in units of the inter-site spacing, $a$, equal to half the wavelength of
the lasers defining the optical lattice. 

\begin{figure}
\includegraphics[clip,width=3.0in]{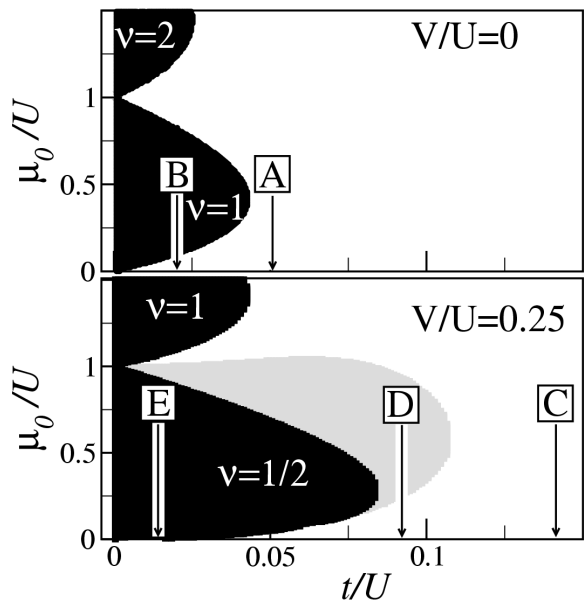}
\caption{ Mean-field phase diagram of the Bose-Hubbard 
(extended Bose-Hubbard) model in the top (bottom) panel showing
chemical potential versus hopping in an infinite system.  The 
Mott and density wave phases are indicated by black 
at integer and half filling, $\nu$, respectively.  The 
grey region indicates supersolid and the remaining
white region superfluid.  The points A through E label the following values
of ($t/U$, $\mu_0 /U$): (0.05,0.6), (0.02,0.6), (0.14,0.8), (0.09,0.8), 
and (0.015,0.8).  The arrows show the chemical potential shifts due to
an external trapping potential.  
}
\label{fig1}
\end{figure}

To study the experimentally observable ground state properties 
of Eq.~(\ref{Hex}) we
first consider the Gutzwiller, mean-field phase diagram in the absence
of a trapping potential, $\kappa=0$ \cite{Fisher}.  We obtain 
the phase digram using a variational state \cite{Rokhsar,Jaksch} in the 
Fock number basis, $\vert N_i \rangle$:   
$ 
\vert \psi_i \rangle =\sum_{N_i=0}^{N_m-1} f^{i}_{N_i} \vert N_i\rangle, 
$
where we fix the $SN_m$ variational parameters $f^{i}_{N_i}$ on
$S$ sublattices.  We find that $N_m=6$ ensures convergence in the 
calculations shown here.
Evaluating  
$
\text{min}\left[\langle \psi_{i\in S} 
\vert H \vert \psi_{i\in S} \rangle \right]
$ 
yields the phase diagram shown in Fig.~\ref{fig1}. 
 In the top panel we choose $V=0$ 
giving only two phases:  the Mott phase 
and the superfluid phase.
The incompressible Mott phase occurs at integer filling, $\nu$, with
order parameter $\langle n_i \rangle\in$integer while
the superfluid can have variable filling with order parameter $\langle
b_i \rangle \neq 0$.  
The bottom panel shows the
mean-field phase diagram with $V=U/z$, where $z=4$ is the number of
nearest neighbors.  Two new phases appear:  the incompressible density
wave phase at $\nu =1/2$ and the supersolid phase.  The  
density wave phase on a square lattice has  
checkerboard oscillations in density with order parameter 
$
(-1)^{(i_x+i_y)}(\langle n_i \rangle-\nu)
$.  
The supersolid consists of
coexisting density wave and superfluid order.  Monte Carlo studies
suggest that the delicate supersolid phase successfully competes with phase
separation \cite{Batrouni} for $V\gtrsim U/z$.  We seek distinct experimental
indicators of density wave and supersolid order 
as captured by a zero temperature, modified Gutzwiller mean-field theory 
in a model which includes the trapping potential.  The technique
employed here can be extended to include finite temperature effects.
In this work we focus on zero temperature properties as
a starting point.   

Inclusion of the trapping potential mixes phases within the trap and
may obscure experimental signatures of the phases
discussed above.  The center of the trap corresponds to a point in 
the phase diagrams of Fig.~\ref{fig1}.  The arrows indicate   
positions moving towards the edge of the
trap which see an effectively lower chemical potential thereby mixing
phases within the trap.  We incorporate mixing by modifying the
Gutzwiller ansatz to bring out an important realistic feature; 
{\em smooth} modulation of the density throughout the trap.  We allow 
the variational parameters  $f^{i}_{N_i}$
to vary from site to site while minimizing the energy of the total
Hamiltonian:      
$
\text{min}\left[\langle \Psi \vert H \vert \Psi  \rangle \right],
$
where $\vert \Psi \rangle=\prod_{i} \vert \psi_i \rangle$ generalizes
$\vert \psi_i \rangle$.
This leaves a
minimization problem over $N_s N_m $ variables, where $N_s$ is the total 
number of sites.  As an initial configuration we choose the 
output from our initial mean-field 
minimization performed independently 
at each site.  First order correlation
functions calculated with this two step procedure show \cite{Zakrzewski} 
close agreement with Monte Carlo studies \cite{Svistunov} for $V=0$
and away from the phase transition.  This suggests that we retain 
quantitative accuracy for $V\neq0$ although the  
techniques employed here remain untested against small system 
Monte Carlo results for $V\neq0$. 
In what follows we consider system sizes
comparable to experiment $N_s=101\times101$ with
$\kappa/U=7\times10^{-4}$ yielding a total number of particles 
$N\sim 10^3$ for $\mu_0/U=0.6-0.8$ and 
$N =\sum_{i}\langle \Psi \vert n_i \vert \Psi \rangle$. 
        
Time of flight absorption imaging of atoms released from trapped
optical lattices provides a direct probe of first {\em and} second
order correlation functions of the initially trapped state.  We 
first consider the momentum distribution 
function measured through an average of images of 
expanding atomic clouds after a time $T$.  Assuming that the 
particles interact weakly after the trap is adiabatically turned off, 
the number of particles at
position $\bm{r}$ in the expanding cloud is given by \cite{Svistunov}:     
$
\langle n(\bm{r}) \rangle_{T}\approx\vert \tilde{w}(\bm{Q}(\bm{r})) \vert ^2 
\rho(\bm{Q}(\bm{r})),
$
where $\bm{Q}(\bm{r})=am\bm{r}/\hbar T$ for particles of mass $m$.
The angular brackets indicate averaging, which, experimentally, occurs 
over several shots.  $\tilde{w}$ is the Fourier transform of 
the Wannier function.  For 
illustration we take particles in the lowest band of a deep lattice.
The harmonic approximation then gives 
(using the conventions of Ref.~\cite{Scarola}):
$
\vert \tilde{w}(\bm{Q})\vert^2
\approx \vert\tilde{w}(0) \vert^2\exp[-\vert\bm{Q}\vert^2/(\pi^2\sqrt{V_L})],
$
for a lattice depth $V_L$ in units of the photon recoil 
energy.  $\rho$ is a first order correlation function 
related to the distribution of lattice momenta $\bm{k}$ in the
trapped system:
\begin{eqnarray}
\rho(\bm{k})=
\sum_{i,j}
e^{i(\bm{R}_i-\bm{R}_j)\cdot\bm{k}}\langle 
b^{\dagger}_{i}b^{\vphantom{\dagger}}_{j}\rangle.
\label{rho}
\end{eqnarray}
Peaks in $\rho$ indicate inter-site phase coherence.  From
shot-to-shot a coherent state maintains, to a good approximation, 
the same phase relationship between Fourier components in the
averaging process.  However, for incoherent states with random Fourier 
components, i.e. the Mott and density wave phases, the 
averaging process eliminates any peak in the Fourier spectrum.  One
can overcome this loss of information by averaging the
second order, noise correlation function. 
  
Noise in the imaged atomic momentum distribution provides a direct
measure of higher order correlation functions.  Shot-to-shot averaging
in the quantity
$
\langle n(\bm{r})n(\bm{r}') \rangle_T
-\langle n(\bm{r}) \rangle_T \langle n(\bm{r}') \rangle_T
$ 
 yields results proportional to momentum
correlations in the ground state of the trapped system \cite{Altman}:    
\begin{eqnarray}
\mathcal{G}(\bm{r},\bm{r}') \sim 
\vert \tilde{w}(\bm{Q}(\bm{r})) \vert ^2 
\vert \tilde{w}(\bm{Q}(\bm{r}'))\vert ^2 
G(\bm{Q}(\bm{r}),\bm{Q}(\bm{r}'))
\nonumber
\end{eqnarray}
and, in contrast to the first order correlation function, 
does not vanish in finite sized system measurements of incoherent states.  
Higher order terms arise in the normalized, second order correlation
function, $\overline{G}(\bm{k},\bm{k}')$, defined as:
\begin{eqnarray}
\frac{G(\bm{k},\bm{k'})}{\rho(\bm{k})\rho(\bm{k}')}&=&
\frac{\sum_{i i' j j'}e^{i(\bm{R}_{ii'}\cdot\bm{k}+\bm{R}_{jj'}\cdot\bm{k}')}
\langle b^{\dagger}_{i}b^{\dagger}_{j}
b^{\vphantom{\dagger}}_{j'}b^{\vphantom{\dagger}}_{i'}\rangle}
{\rho(\bm{k})\rho(\bm{k}')}
-1
\nonumber
\\
&+&\left(\frac{ma}{\hbar T}\right)
\frac{\delta(\bm{k}-\bm{k}')}{\vert \tilde{w}(\bm{k}')\vert ^2 \rho(\bm{k}')},
\end{eqnarray}
providing a direct
measure of Mott and density wave order.  
The delta function results from normal ordering \cite{Altman}.  
We evaluate $\rho$ and $G$ 
using the output of our modified Gutzwiller mean-field theory, 
$\vert \Psi \rangle$.  


In a translationally invariant system (i.e. $\kappa=0$) 
$\rho$ and $G$ qualitatively distinguish the 
ground states of Eq.~(\ref{Hex}).   Macroscopic 
occupation of a Bloch mode of the optical lattice at the 
reciprocal lattice vectors 
should lead to peaks in $\rho$ demonstrating a 
spontaneously broken $U(1)$ symmetry 
associated with the global phase, and therefore superfluid order.  If 
the peak exists at wavevectors other than the reciprocal 
lattice vectors we expect, in addition, spontaneously broken 
translational symmetry, i.e. supersolid order.  
(In contrast, applied potentials with spatial \cite{Phillips2} and 
temporal \cite{Chu} modulation of the the host lattice have led 
to the observation of coherence peaks away from 
the reciprocal lattice vectors not
associated with a supersolid.)  
Furthermore, strong peaks in
$\overline{G}$ at the reciprocal lattice vectors of a translationally invariant 
optical lattice indicate Mott order while strong peaks in $\overline{G}$ at half
the reciprocal lattice vector indicate spontaneously broken 
translational symmetry and therefore density wave order.  As we will
see both correlation functions, in a trapped system,   
merely quantitatively distinguish between the supersolid and 
other phases.

\begin{figure}
\includegraphics[clip,width=3.0in]{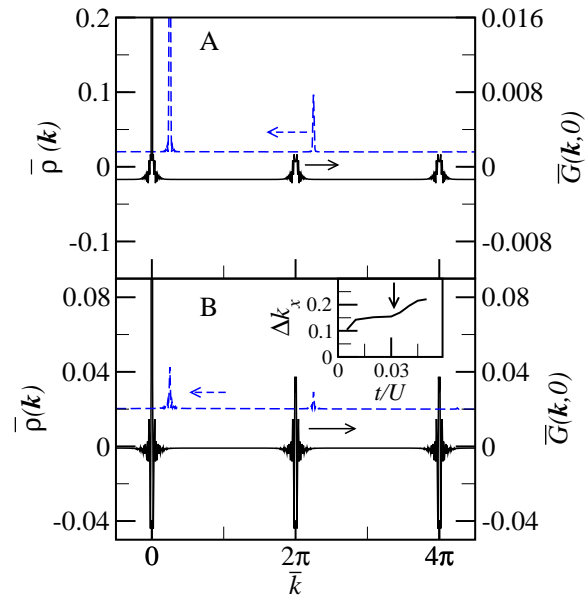}
\caption{  The first order, $\overline{\rho}$, (left $y$-axis, dashed
  line) and second order, $\overline{G}$, (right $y$-axis, solid line) 
normalized correlation functions plotted as a function of wavevector,
$\overline{k}$, for the trapped Bose-Hubbard model.  
We set $\overline{k}=k_x=k_y$.  The dashed lines are shifted up 
by 0.02 and right by $\pi/4$ for clarity.
The superfluid phase (top panel, point A in
Fig.~\ref{fig1}) shows strong peaks  
in $\overline{\rho}$ at the reciprocal lattice 
vectors $\bm{K}$.  In this and in the
following figures we choose a deep optical lattice, $V_L=20$, which 
determines the weight of $\overline{\rho}$ at higher $\overline{k}$.  The 
Mott phase (bottom panel, point B in Fig.~\ref{fig1}) shows strong
peaks in $\overline{G}$ at the reciprocal lattice vectors $\bm{K}$.
The inset plots the full width at half maximum of the central peak in 
$\overline{G}$ at $\bm{k}=(2\pi,2\pi)$ versus hopping for $\mu/U=0.7$.    
The vertical arrow indicates the position of a superfluid to Mott transition 
at the trap center.
}
\label{fig2}
\end{figure}

We first verify that, as expected, superfluid and Mott 
order in the Bose-Hubbard model remain identifiable through
first and second order correlation functions in a trapped 
system.  Fig.~\ref{fig2} plots
both correlation functions for $\kappa/U=7\times10^{-4}$ at
parameters corresponding to the points A (top panel) and B (bottom panel) of
Fig.~\ref{fig1} as a function of $\overline{k}$.  
The left $y$-axis plots the
normalized momentum distribution function,
$
\overline{\rho}(\bm{k})\equiv\rho(\bm{k})
\vert \tilde{w}(\bm{k})\vert ^2 /\vert N \tilde{w}(0)\vert ^2, 
$
 while the right $y$-axis plots     
$
\overline{G}(\bm{k},0).
$ 
  The superfluid phase (top panel) shows a strong coherence peak at the
reciprocal lattice vectors  $\bm{K}\equiv (2\pi j,2\pi j)$, for
integer $j$,
($\overline{\rho}(2\pi,2\pi)/\overline{\rho}(0,0) \sim 0.3$) with little 
signal in the noise.  The exponential decay of $\tilde{w}(\bm{k})$ strongly
suppresses the peak in $\overline{\rho}$ at $(2\pi,2\pi)$ as opposed to the peak at
$(2\pi,0)$ (not shown).  The Mott
phase (bottom panel) is identifiable through a 
strong peak in $\overline{G}$, at 
the reciprocal lattice vectors.  Note that here there is 
still a small amount of superfluid at the edge of the 
system, identifiable through weak peaks in $\overline{\rho}$.  

Individual noise correlation peaks 
reveal position in the phase diagram through an additional feature, 
peak width. 
Increasing $t/U$ through the Mott phase we find the 
noise correlation peaks to decrease
in height and split.  The splitting results from our
choice of normalization.  Other suitably chosen normalization schemes 
can eliminate the splitting by excluding coherence peaks from the definition 
of the normalization \cite{Spielman}.  In addition, we predict 
that the widths of the peaks vary inversely with the size of the Mott 
domains (rather than the total system size), in qualitative 
agreement with recent experimental results \cite{Spielman}.  The inset to 
Fig.~\ref{fig2} shows the full width at half 
maximum along the $k_x$ axis, $\Delta k_x$, of 
the central $(2\pi,2\pi)$ noise correlation peak increasing 
with $t/U$.  An abrupt increase in peak width 
occurs when the center of the trap crosses from the Mott to the 
superfluid regime.  We conclude that noise correlation peak widths yield
detailed information related to the location of the system in the
phase diagram.  

\begin{figure}
\includegraphics[clip,width=3.0in]{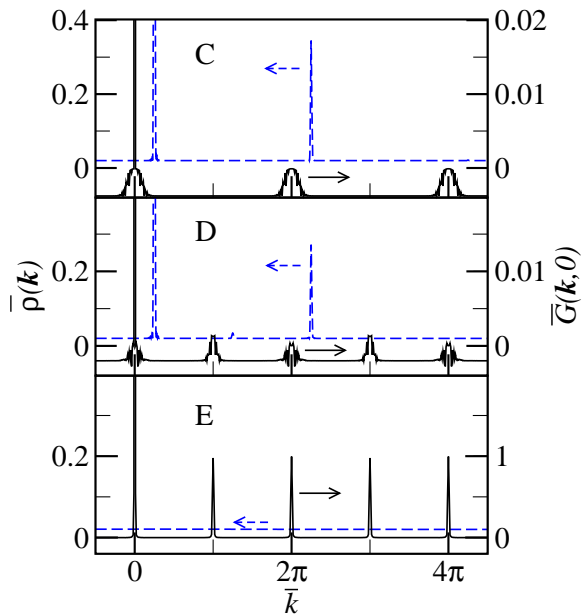}
\caption{  The same as Fig.~\ref{fig2} but for the points: C (top 
panel, superfluid), D (middle panel, supersolid), E (bottom panel,
density wave) of Fig.~\ref{fig1}.  
}
\label{fig3}
\end{figure}

Fig.~\ref{fig3} shows correlation functions of the extended 
Bose-Hubbard model ($V/U=0.25$) along 
the points C (top panel), D (middle panel), and E (bottom
panel) of Fig.~\ref{fig1}.  
The top panel shows signatures of
superfluidity; a strong peak in  $\overline{\rho}$, with  
$\overline{\rho}(2\pi,2\pi)/\overline{\rho}(0,0) \sim 0.18$, 
and weak structure
in $\overline{G}$.  
The bottom panel, deep in the density wave regime, 
shows signatures of checkerboard density wave order only; strong peaks in noise
correlations at $\bm{k}=\bm{K}$ {\em and} $\bm{K}/2$.  The paucity of
superfluid at the edge yields almost no structure in $\overline{\rho}$.
The middle panel is plotted for parameters with supersolid at the trap
center and superfluid near the edge.  Here we see 
a coherence peak at the reciprocal lattice vectors $\bm{K}$ and very
weak structure in $\overline{G}$.  There is also a 
very weak peak ($\sim 0.01$) 
in $\overline{\rho}$ at $(\pi, \pi)$.  We conclude that here weak
supersolid order is nearly indistinguishable from superfluid order.

The supersolid shows a small signal in our mean-field estimates of the 
first order correlation function.  We define the strength of a 
supersolid phenomenologically through $\rho(\pi,\pi)/\rho(0,0)$ which
is sensitive to spatial modulation of the superfluid order parameter, 
$\langle b_i \rangle$.
The nearby superfluid and density wave phases surrounding the supersolid
phase in parameter space show no modulation 
in $\langle b_i \rangle$.  We then expect
the supersolid to exhibit only weak structure in $\langle b_i
\rangle$ and therefore $\rho(\pi,\pi)$.  The modulation amplitude can be
enhanced if the density wave-supersolid and superfluid-supersolid
phase boundaries lie far apart in parameter space.  We may therefore reach the
strong supersolid regime by increasing $V/U$.

\begin{figure}
\includegraphics[clip,width=3.0in]{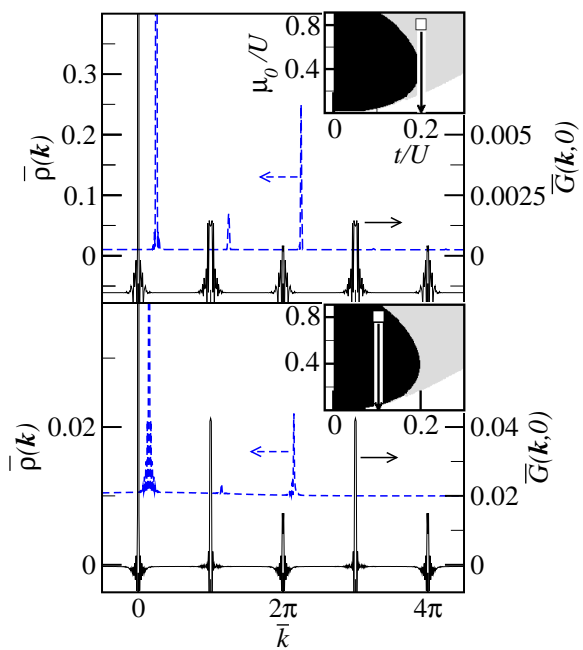}
\caption{  The main panels plot the same as Fig.~\ref{fig2} 
but for $V/U=1$ and $\mu/U=0.8$ with the dashed lines shifted up 
by 0.01 and right by $\pi/4$ for clarity.  The main top (bottom)
panel is in the supersolid (density wave) regime with $t/U=0.2$
($t/U=0.1$).  The insets show the phase diagram (as in
Fig.~\ref{fig1} but with $V/U=1$) near the $\nu=1/2$ density wave.  
}
\label{fig4}
\end{figure}

We now consider the strong supersolid regime in a trap.  The main panels 
of Fig.~\ref{fig4} plot the same as Fig.~\ref{fig2} but for $V/U=1$,
$\mu/U=0.8$, $t/U=0.2$ (top panel), and $t/U=0.1$ (bottom panel).  
We choose the parameters
for the top panel so that the supersolid is strongest at
the center of the trap with superfluid near the edge as indicated in
the inset.  We again see a strong, first order coherence peak at
$\bm{k}=\bm{K}$ with weak structure in $\overline{G}$.  We 
also find $\overline{\rho}(\pi,\pi)/\overline{\rho}(0,0)\sim 0.03$.

We now compare the correlation functions of a strong supersolid with
those of the trapped density wave.  The bottom panel of Fig.~\ref{fig4} lies
in the density wave regime with a  
significant amount of superfluid near the edge.  
A strong enhancement in noise correlations at  
$\bm{K}/2$ signals the onset of density wave order.  
Superfluidity near the edge enhances $\overline{\rho}(\bm{K})$ and
splits the $\overline{G}(\bm{K},0)$ peaks.  There is also a 
small but surprising feature in $\overline{\rho}$ near $(\pi,\pi)$.  The 
weak $(\pi,\pi)$ feature, 
$\overline{\rho}(\pi,\pi)/\overline{\rho}(0,0)\sim 0.02$, previously 
associated with the supersolid phase, arises here as the
density wave and trap corrugate the superfluid at the edge.  The modulated
superfluid does not spontaneously break translational symmetry 
along the edge but consists of a mixture of phase boundaries mitigated by the
trapping potential.  We further find that 
the edge mode density oscillations never reach zero, as in the case of a
bulk supersolid.  The
essential differences in going from the top panel (comprising a trapped
supersolid) to the bottom panel (a trapped density wave) are  
therefore merely quantitative; a large enhancement in
$\overline{G}(\bm{K}/2,0)$ 
combined with a sharp drop in $\overline{\rho}(\bm{K}/2)$ 
and $\overline{\rho}(\bm{K})$.

Our interesting and important findings are that (1)  noise
correlations enable relatively precise determination of phase
boundaries in trapped systems and are therefore 
fundamental in providing unambiguous observation of
the supersolid phase, and (2) the edge of the cold atom density wave
may accommodate a superfluid component, mimicking the existence
of an edge supersolid.
 
We thank I.B. Spielman and J.V. Porto for valuable discussions.  VS
and SDS acknowledge support from ARO-ARDA and NSA-LPS.




\begin{thebibliography}{}

\bibitem{Phillips} P. Verkerk {\em et al.}, 
Phys. Rev. Lett. \textbf{68}, 3861 (1992);
P. S. Jessen {\em et al.}, 
Phys. Rev. Lett. \textbf{69}, 49 (1992);  
A. Hemmerich and T. W. Hansch, 
Phys. Rev. Lett. \textbf{70}, 410 (1993).

\bibitem{Jaksch} D. Jaksch {\em et al.},
Phys. Rev. Lett. \textbf{81}, 3108 (1998).

\bibitem{Greiner} M. Greiner {\em et al.},
Nature \textbf{415}, 39 (2002).

\bibitem{Folling} S. F\"{o}lling {\em et al.}, 
Nature \textbf{434}, 481 (2005).

\bibitem{Duan} L.M. Duan {\em et al.}, 
Phys. Rev. Lett. \textbf{91}, 090402 (2003). 

\bibitem{Girvin} A. Isacsson and S.M. Girvin,
Phys. Rev. A \textbf{72}, 053604 (2005).

\bibitem{Sorensen} A.S. S\o rensen {\em et al.},
Phys. Rev. Lett. \textbf{94}, 086803 (2005).

\bibitem{Scarola} V.W. Scarola and S. Das Sarma,
Phys. Rev. Lett. \textbf{95}, 033003 (2005).

\bibitem{Altman} E. Altman {\em et al.},
Phys. Rev. A \textbf{70}, 013603 (2004).

\bibitem{Hadzibabic} Z. Hadzibabic {\em et al.},
Phys. Rev. Lett. \textbf{93}, 180403 (2004). 

\bibitem{Greiner2} M. Greiner {\em et al.},
Phys. Rev. Lett. \textbf{94}, 110401 (2005).

\bibitem{Goral} K. Goral {\em et al.}, 
Phys. Rev. Lett. \textbf{88}, 170406 (2002).

\bibitem{Axel} A. Griesmaier {\em et al.}, 
Phys. Rev. Lett. \textbf{94}, 160401 (2005). 

\bibitem{Fisher} M.P.A. Fisher {\em et al.},
Phys. Rev. B. \textbf{40}, 546 (1989).

\bibitem{Rokhsar} D.S. Rokhsar and B.G. Kotliar, 
Phys. Rev. B \textbf{44}, 10328 (1991).

\bibitem{Batrouni} G.G. Batrouni {\em et al.},
Phys. Rev. Lett. \textbf{84}, 1599 (2000); 
P. Sengupta {\em et al.}, 
Phys. Rev. Lett. \textbf{94}, 207202 (2005).

\bibitem{Zakrzewski} J. Zakrzewski,
Phys. Rev. A \textbf{71}, 043601 (2005).

\bibitem{Svistunov} V. A. Kashurnikov {\em et al.},
Phys. Rev. A \textbf{66}, 031601(R) (2002).

\bibitem{Phillips2} S. Peil {\em et al.},
Phys. Rev. A \textbf{67}, 051603(R) (2003).

\bibitem{Chu} N. Gemelke {\em et al.},
Phys. Rev. Lett. \textbf{95}, 170404 (2005).

\bibitem{Spielman} I.B. Spielman and J.V. Porto to be published; 
(private communication).

\end{thebibliography}
\end{document}